# The Spectrum of the 4-Generation Dirac-Kähler Extension of the SM

Alexander N. Jourjine[1]

**Abstract**

We compute the mass spectrum of the fermionic sector of the Dirac-Kähler extension of the SM (DK-SM) by showing that there exists a Bogoliubov transformation that transforms the DK-SM into a flavor U(4) extension of the SM (SM-4) with a particular choice of masses and mixing textures. Mass relations of the model allow determination of masses of the 4th generation. Tree level prediction for the mass of the 4$^{th}$ charged lepton is 370 GeV. The model selects the normal hierarchy for neutrino masses and reproduces naturally the near tri-bimaximal and quark mixing textures. The electron neutrino and the 4th neutrino masses are related via a see-saw-like mechanism.

## 1. Introduction

Originally Dirac-Kähler (DK) spinors became interesting within the context of lattice fermion doubling problem [1, 2]. These objects, which describe multiple fermionic degrees of freedom, do not exhibit fermion doubling and their naïve lattice limit coincides with their continuum analog. It was then realized that the extra degrees of freedom could be useful to describe the generation structure of the SM [3, 4]. However, because of a number of technical issues [5, 6] and the general belief that the fourth generation was ruled out by the EW data, this approach to solving the generation puzzle was not pursued.

It is now clear that four-generation extensions do not necessarily contradict the data [7-11]. Moreover, some of the technical problems that prevented the use of DK spinors in realistic models have been recently solved [12]. In this Letter, which should be considered as a companion article to [12] and to which the reader is referred for details on DK spinors, we describe the DK-SM, an extension of the SM to a four generation model that uses DK spinors, and show that there exists a Bogoliubov transformation in the phase space that maps its Fock space into that of a SM-4 with a particular choice of masses and mixing. Its main two differences from other four-generation extensions come from flipped anti-commutation relations for the 3$^d$ and 4$^{th}$ generations and from the form of the mass matrices. In the DK-SM they are not arbitrary but belong to $U(2,2)$. As a result, the form of its mixing is constrained and the lepto-quark sector of the model contains four mass relations, allowing determination of the masses of the 4$^{th}$ generation, if the masses of the first three generations are known. In this Letter we compute the mass spectrum of the theory and compare tree-level predictions for the 4$^{th}$ generation masses with their experimentally known bounds. Mixing in the DK-SM shall be mentioned only briefly. We describe the textures of the DK-SM mixing matrices that apply both to lepton and quark sectors. Details on mixing shall be presented elsewhere.

The Letter is organized as follows. In Section 2 we describe DK spinors and the DK-SM. In Section 3 we compute its mass spectrum. In Section 4 we compare the tree-level predictions for masses of lepto-quarks with the experimental data and discuss mixing. Section 5 is a summary.

---

[1] E-mail: jalnich@gmx.net. Address for correspondence: Postfach 53 01 11, 01291 Dresden, Germany.

## 2. Spinbeins and the DK Extension of the SM

Given a pseudo-Riemannian manifold $M$, a DK spinor is defined as an inhomogeneous differential form $f$ with values in the Lie algebra of the internal symmetry group [13, 14]. One defines massless on-shell DK spinors coupled to background gravity as solutions of $(d - \delta)f = 0$, where $d$ is the exterior derivative, and $-\delta$ is its adjoint. Unlike the Dirac spinor case, the coupling of DK spinors to gravity is unique. The relation between differential forms and spinors on $M$ can be seen in the special basis in the space of differential forms [2]. Given an orthonormal frame one-forms $e^a$ and tangent space $\gamma$-matrices $\gamma^a$, it is defined by $Z = \sum (1/p!) \gamma_{a_p} \cdots \gamma_{a_1} e^{a_1} \cdots e^{a_p}$ so that $f = Tr(\Psi Z)$. Under the general coordinate transformations the coefficient matrix $\Psi$ transforms as a scalar, while under local Lorentz frame rotations $e^a \to \Lambda^a_b e^b$ it transforms as a multiplet of integer-valued spin fields according to $\Psi(x) \to S(\Lambda) \Psi(x) S(\Lambda)^{-1}$, where $S$ denotes a spinor representation of $SO(1,3)$. When $M$ is flat then the equation on $\Psi$ that is obtained from $(d-\delta)f = 0$ reduces to four Dirac equations and one can identify the four columns of $\Psi$ with four generation of Dirac spinors. However, such identification is possible only for flat $M$ [14].

A generally covariant way of identifying Dirac degrees of freedom contained in a DK spinor was described in [12]. One takes two sets of Dirac spinors $\xi^A$ and $\eta^A$, $A = 1, \ldots, 4$, and defines the spinbein decomposition of $\Psi$ by

$$\Psi_{\alpha\beta} = \xi^A_\alpha \overline{\overline{\eta}}^A_\beta, \tag{1}$$

where the commuting dimensionless spinbein $\eta^A$ satisfies the orthonormality conditions

$$\overline{\overline{\eta}}^A_\alpha \eta^B_\alpha = \delta^{AB}, \quad \eta^A_\alpha \overline{\overline{\eta}}^A_\beta = \delta_{\alpha\beta}. \tag{2}$$

Here $\overline{\overline{\eta}}^A$ denotes the DK conjugate of $\eta^A$, $\overline{\overline{\eta}}^A = \Gamma^{AB} \overline{\eta}^B$, $\overline{\eta}^B = \eta^{B+} \gamma^0$, $\Gamma \equiv diag(1,1,-1,-1)$. An example of a spinbein that is often used is $u^{(A)}(k) = \{u^{(\alpha)}(k), v^{(\beta)}(k), \alpha = 1,2, \beta = 1,2\}$, where $u^{(\alpha)}(k), v^{(\beta)}(k)$ form the set of normalized positive and negative energy solutions of the Dirac equations in the phase space. Here the role of the generation index is played by the two combined spin indexes.

Representation (1, 2) is trivially invariant with respect to simultaneous local $U(2,2)$ transformations of $\xi^A$ and $\eta^A$: $\xi^A \to V^{AB} \xi^B$, $\eta^A \to V^{AB} \eta^B$, $\Gamma V^+ \Gamma V = 1$. Therefore, for Minkowski $M$ the coordinate dependence of $\eta^A$ can be eliminated globally and in (1) we can choose a constant spinbein $\eta^A = const$. This condition fixes the $U(2,2)$ gauge, which is called the unitary gauge. Choosing the unitary gauge breaks local $U(2,2)$ and transfers all physical degrees of freedom of $\Psi$ to the multiplet of four Dirac spinors $\xi^A$.

We now describe the fermionic sector of the DK-SM. In the SM the left- and the right-handed lepto-quarks transform in different representations of $SU(2)$. As a result, dimension three gauge invariant mass terms for chiral fermions are not allowed. In the DK-SM they are allowed. Consider three DK $SU(2)$ doublet fields $\Psi$, $\Phi_u$, $\Phi_d$ with spinbein decompositions that use an $SU(2)$ singlet spinbein $\eta$ and two factorizable $SU(2)$ doublet spinbeins $\chi$ and $\theta$

$$\Psi = Q\bar{\bar{\eta}}, \qquad \Phi_u = u\bar{\bar{\chi}}, \qquad \Phi_d = d\bar{\bar{\theta}}. \tag{3}$$

$$\chi_i = \varphi_i \hat{\chi}, \qquad \bar{\bar{\chi}}\hat{\chi} = 1, \qquad \theta_i = \varphi_i^* \hat{\theta}, \qquad \bar{\bar{\hat{\theta}}}\hat{\theta} = 1, \qquad \varphi_i^* \varphi_i = 1. \tag{4}$$

The choice of factorizable spinbeins insures that they are non-dynamical [12]. Here $\varphi_i$ is a $SU(2)$ doublet of scalars.

The left-handed $Q^A$ and the right-handed $u^A, d^A$ quark fields represent the four generations of Dirac spinor particles transforming in the fundamental representation of $SU(3)$. Additionally, we shall take $u, d, \eta$ as $SU(2)$ singlets while $Q, \chi, \theta$ are taken to be $SU(2)$ doublets with gauge transforms

$$Q \to TQ, \; \chi \to T\chi, \; \theta \to T^*\theta, \qquad T, T^* \in SU(2), \tag{5}$$

so that under $SU(2)$ the three DK fields transform as doublets

$$\Psi \to T\Psi, \Phi_u \to T^*\Phi_u, \Phi_d \to T\Phi_d. \tag{6}$$

If we assume that $Q, u, d$ are $SU(3)$ singlets then the following discussion describes four generations of leptons with Dirac mass term for neutrinos.

The quark sector of the DK-SM is described by the $SU(2)$ invariant Lagrangian with three chiral DK fields given by (3)

$$\begin{aligned}\mathcal{L} = Tr\Big[&\bar{\bar{\Psi}}(i\mathbb{D})\Psi + \bar{\bar{\Phi}}_u(i\mathbb{D})\Phi_u + \bar{\bar{\Phi}}_d(i\mathbb{D})\Phi_d\Big] - \\ &- m_u Tr\Big[\bar{\bar{\Psi}}(i\sigma^2)\Phi_u - \bar{\bar{\Phi}}_u(i\sigma^2)\Psi\Big] - m_d Tr\Big[\bar{\bar{\Psi}}\Phi_d + \bar{\bar{\Phi}}_d\Psi\Big],\end{aligned} \tag{7}$$

where $\mathbb{D}$ are the usual $G = SU_C(3) \times SU_L(2) \times U_Y(1)$ gauged derivatives when acting on the Dirac spinor components of the DK fields and are ungauged derivatives when acting on spinbeins [12]. Parameters $m_u, m_d$ are bare masses for the up and down quarks. The lepton sector is obtained by dropping $SU_C(3)$ from $G$. The remaining terms in the DK-SM are the same as in the SM, except that adding Higgs-related terms is optional. Such terms can be used to generate the gauge field mass but in principle it can be generated using the DK-SM surrogate Higgs fields constructed from contractions of the three spinbeins [12]. In the unitary gauge with $\eta = const$, $\hat{\chi} = const$, $\hat{\theta} = const$, $\varphi = const$ we obtain

$$\begin{aligned}\mathcal{L} &= \mathcal{L}_\mathcal{K} + \mathcal{L}_\mathcal{M}, \\ \mathcal{L}_\mathcal{K} &= \bar{\bar{Q}}^A_i(i\mathbb{D})Q^A_i + \bar{\bar{u}}^A(i\mathbb{D})u^A + \bar{\bar{d}}^A(i\mathbb{D})d^A, \\ \mathcal{L}_\mathcal{M} &= -m_u\Big(\mathcal{M}_u^{AB}\bar{\bar{Q}}^A_i(i\sigma^2)\varphi_i^* u^B - \tilde{\mathcal{M}}_u^{AB}\bar{\bar{u}}^B(i\sigma^2)\varphi_i Q^A_i\Big) - m_d\Big(\mathcal{M}_d^{AB}\bar{\bar{Q}}^A_i\varphi_i d^B + \tilde{\mathcal{M}}_d^{AB}\bar{\bar{d}}^A\varphi_i^* Q^B_i\Big),\end{aligned} \tag{8}$$

$$\mathcal{M}_u^{AB} = \bar{\bar{\hat{\chi}}}^B \eta^A, \qquad \tilde{\mathcal{M}}_u^{AB} = \bar{\bar{\eta}}^B \hat{\chi}^A, \qquad \mathcal{M}_d^{AB} = \bar{\bar{\hat{\theta}}}^B \eta^A, \qquad \tilde{\mathcal{M}}_d^{AB} = \bar{\bar{\eta}}^B \hat{\theta}^A. \tag{9}$$

In general $\hat{\chi} \neq \hat{\theta}$ and $\mathcal{M}_u^{AB}, \mathcal{M}_d^{AB}$ are independent from each other. In (8) the non-dynamical field $\varphi$ plays the role of the surrogate Higgs field of the SM. Combined with mass matrices $m_u \mathcal{M}_u^{AB}$, $m_d \mathcal{M}_d^{AB}$ it generates the familiar Yukawa terms for massless up and down quarks. If we choose $\varphi_1 = 0$, $\varphi_2 = 1$ then (8) reduces to the action for four generations of two multiplets of massive Dirac spinors with masses $m_u^A$, $m_d^A$, $A = 1,\ldots,4$ that have to be determined by diagonalizing the mass terms of

$$\mathcal{L} = \overline{\overline{Q}}_i^A (i\!\!\not{D}) Q_i^A + \overline{\overline{u}}^A (i\!\!\not{D}) u^A + \overline{\overline{d}}^A (i\!\!\not{D}) d^A - \left( m_u \overline{\overline{Q}}_1^A \mathcal{M}_u^{AB} u^B + m_d \overline{\overline{Q}}_2^A \mathcal{M}_d^{AB} d^B + c.c. \right). \quad (10)$$

### 3. The Mass Spectrum of the DK-SM

Spinbein orthonormality (2) implies $\mathcal{M}_{u,d} \in U(2,2)$. On classical level we can use invariance of $\mathcal{L}_K$ with respect to global $U(2,2)$ to absorb $\mathcal{M}_{u,d}$ in (10) in the definitions of $u$, $d$ and, therefore, classically the fermion masses are generation-independent $m_u^A = m_u$, $m_d^A = m_d$. The mass spectrum of the classical DK-SM is completely degenerate. Quantized DK-SM has a generally non-degenerate spectrum, for quantization breaks global $U(2,2)$ to $U(2) \times U(2)$ and, hence, the matrices $\mathcal{M}_u^{AB}$, $\mathcal{M}_d^{AB}$ in (10) can no longer be absorbed in the definitions of $u$, $d$ [12]. As a result, quantization lifts $U(2,2)$ mass degeneracy and generates non-trivial mixing matrices for the quark and, independently, for lepton sectors.

We now compute the mass spectrum of the DK-SM theory. Mixing in the DK-SM shall be described in a follow-up article. Since derivation of the spectrum of the DK-SM contains some unusual features, for clarity we carry out the procedure in parallel both for SM-4, the naïve extension of the SM, where the kinetic part of the action has global $U(4)$ flavor symmetry, and for DK-SM. Mass matrix diagonalization of DK action differs from its SM-4 analog. In the latter case one can diagonalize by global $U(4)$ flavor transformation of the Dirac fields, similarly to how it is done in the SM. Such transformation does not exist for the DK-SM action, because its kinetic part is not $U(4)$ invariant. The reason for the difference can also be seen in the phase space. Because of the flipped anti-commutation relations for the generations 3 and 4 in (15, 16) below, some of the mass terms of the DK-SM action contribute to its kinetic component.

As a preliminary step we describe a version of the Cartan group element decomposition for $U(2,2)$. It is analogous to that for $M \in GL(4)$, which is used to diagonalize mass terms in the SM-4 case. Recall that an arbitrary non-degenerate matrix $M \in GL(4)$ can be represented as a product of three matrices, two of which are unitary, while the middle factor is positive and diagonal

$$M = U R_D V, \qquad U, V \in U(4), \ R_D = diag(\lambda_j), \ j = 1,\ldots,4, \ \lambda_k > 0. \quad (11)$$

Similar decomposition for $\mathcal{M} \in U(2,2)$ is given by

$$\mathcal{M} = U R V, \quad (12)$$

where $U, V \in U(2) \times U(2)$ and positive middle factor is given by

$$R = \begin{pmatrix} C & S \\ S & C \end{pmatrix}, \ S = diag(sh\lambda_1, sh\lambda_2), \ C = diag(ch\lambda_1, ch\lambda_2), \quad (13)$$

where parameters $\lambda_1, \lambda_2$ may be taken as non-negative real. The eigenvalues of $R$ pair up and form a spectral set $\left(e^{-\lambda_1}, e^{-\lambda_2}, e^{+\lambda_1}, e^{+\lambda_2}\right)$, where, assuming non-degeneracy of the spectrum, to be definite we chose $\lambda_1 > \lambda_2$. Note that both decompositions are defined up to a unitary factor that commutes with the middle factor. Below, to compare DK-SM and SM-4, we shall assume that in (11) $R_D = R$ and associate the spectral set of R with the matrix

$$R_d = diag\left(e^{-\lambda_1}, e^{-\lambda_2}, e^{+\lambda_1}, e^{+\lambda_2}\right). \tag{14}$$

To compute spectrum we shall also need the normalized off-shell Fourier expansions for the chiral DK fields. In two-component form they are given by

$$\xi_L^A(x) = \int \frac{d^4k}{(2\pi)^4}\left(b_2^A(k)\, w_{L-}(k)e^{-ikx} + d_1^{A+}(k)\, w_{L+}(k)e^{ikx}\right),$$
$$\xi_R^A(x) = \int \frac{d^4k}{(2\pi)^4}\left(b_1^A(k)\, w_{R-}(k)e^{-ikx} + d_2^{A+}(k)\, w_{R+}(k)e^{ikx}\right), \quad A = 1, 2, \tag{15}$$

$$\xi_L^A(x) = \int \frac{d^4k}{(2\pi)^4}\left(b_2^{A+}(k)\, w_{L-}(k)e^{-ikx} + d_1^A(k)\, w_{L+}(k)e^{ikx}\right),$$
$$\xi_R^A(x) = \int \frac{d^4k}{(2\pi)^4}\left(b_1^{A+}(k)\, w_{R-}(k)e^{-ikx} + d_2^A(k)\, w_{R+}(k)e^{ikx}\right), \quad A = 3, 4, \tag{16}$$

where $\left(\xi_L^A(x), \xi_R^A(x)\right)$ stand for $\left(Q_1^A(x), u^A(x)\right)$ or $\left(Q_2^A(x), d^A(x)\right)$ in (10). The standard on-shell expansions are recovered if we use $w_{L,R\pm}(k) \propto \theta(k_0)\delta(k^2)$. The four off-shell Dirac spinor amplitudes $w_{L,R\pm}(k)$ are determined essentially uniquely by requiring that the spin off-diagonal terms in the phase space representation of the free action vanish and are given by

$$w_{L-}(k) = (k \cdot \sigma)^{1/2} u, \qquad w_{L+}(k) = (k \cdot \sigma)^{1/2^+} v,$$
$$w_{R-}(k) = (k \cdot \bar\sigma)^{1/2^+} v, \qquad w_{R+}(k) = -(k \cdot \bar\sigma)^{1/2} u, \tag{17}$$

where we use the chiral representation of the $\gamma$-matrices with $\gamma^5 = -\Gamma$ given in [15], $u, v$ are two constant two-component orthonormal spinors and $k \cdot \sigma = k_\mu \sigma^\mu$, $k \cdot \bar\sigma = k_\mu \bar\sigma^\mu$, $\sigma^\mu = (1, \sigma^i)$, $\bar\sigma^\mu = (1, -\sigma^i)$, where $\sigma^i$, $i = 1, 2, 3$, are the Pauli matrices. In the reference frame with $k_1 = k_2 = 0$ we can take for up spin $u = (0,1)^T$ and for down spin $v = (1,0)^T$. The square root of Hermitean matrices $(k \cdot \sigma)$, $(k \cdot \bar\sigma)$ is uniquely defined via positive sign square roots of their eigenvalues. Up to possible phases, the particular choice (17) reduces on-shell to the standard expressions for massless spinors [15]. The off-shell amplitudes satisfy orthogonality conditions that ensure the vanishing in the free action of the terms containing differing spins

$$w_{L-}^+(k)(k \cdot \bar\sigma) w_{L+}(k) = 0, \qquad w_{R-}^+(k)(k \cdot \sigma) w_{R+}(k) = 0,$$
$$w_{L-}^+(k) w_{R-}(k) = 0, \qquad w_{L+}^+(k) w_{R+}(k) = 0. \tag{18}$$

We now proceed with computing the spectrum for the up quarks. The down quarks spectrum can be obtained analogously. Using (15-17) we can write the up quark action terms in (10) as a bilinear form in a 32-dimensional Grassmann space

$$S = S_0 + S_i, \tag{19}$$

where the free $S_0$ and the interaction $S_i$ parts of the action are given by

$$S_0 = \frac{1}{2}\int \frac{d^4k}{(2\pi)^4}\left(C_1^+(k),\ C_2^+(k)\right)\begin{pmatrix} \mathcal{S}_1(k) & 0 \\ 0 & \mathcal{S}_2(k) \end{pmatrix}\begin{pmatrix} C_1(k) \\ C_2(k) \end{pmatrix}, \tag{20}$$

$$S_i = \frac{1}{2}\int \frac{d^4k}{(2\pi)^4}\frac{d^4k'}{(2\pi)^4}C^+(k)\mathcal{I}(k,k')C(k'),\quad C(k) = (C_1(k),\ C_2(k)),$$

where two sixteen-component vectors $C_\alpha$ with definite spin $\alpha = 1,2$ defined by

$$C_\alpha(k) = \left(b_\alpha^A(k), d_\alpha^A(-k), b_\alpha^{A+}(k), d_\alpha^{A+}(-k)\right)^T. \tag{21}$$

For the SM-4 case the integrand in $S_i$ can be written as

$$C^+(k)\mathcal{I}(k,k')\,C(k') = h_{++}(k,k')d_1^A(-k)d_1^{A+}(-k') + h_{--}(k,k')b_2^{A+}(k)b_2^A(k')$$
$$+ h_{+-}(k,k')d_1^A(-k)b_2^A(k') + h_{-+}(k,k')b_2^{A+}(k)d_1^{A+}(-k') + \ldots,$$

where the remaining four terms are obtained by flipping the anti-commuting $d,b$ factors. The factors $h_{\pm\pm}(k,k')$ are linear combinations of the off-shell creation and annihilation operators of the gauge fields, the on-shell versions of which are defined in [16]. Their exact expressions are not needed for our discussion. They become important when mixing is discussed. For the DK-SM case one obtains an analogous expression by switching the spinorial creation and annihilation operators for $A = 3,4$.

Since below we will be dealing with manifestly covariant expressions for various quantities, for simplicity from now on we shall use the in the $k_1 = k_2 = 0$ reference frame. The factors $\mathcal{S}_\alpha(k) = \mathcal{S}_\alpha(k,M)$, where $M = f\,R_d$, satisfy

$$\mathcal{S}_\alpha^+(k,M) = \mathcal{S}_\alpha(k,M),\ \ \mathcal{S}_2(k,M) = \mathcal{S}_1(k,M^+), \tag{22}$$

and for the SM-4 case are defined by

$$\mathcal{S}_1(k) = \begin{pmatrix} g_-I & 0 & 0 & -f^*R_d \\ 0 & g_+I & f^*R_d & 0 \\ 0 & f\,R_d & -g_-I & 0 \\ -f\,R_d & 0 & 0 & -g_+I \end{pmatrix},\ \mathcal{S}_2(k) = \begin{pmatrix} g_-I & 0 & 0 & -f\,R_d \\ 0 & g_+I & f\,R_d & 0 \\ 0 & f^*R_d & -g_-I & 0 \\ -f^*R_d & 0 & 0 & -g_+I \end{pmatrix}, \tag{23}$$

where $I$ is the $4\times 4$ unit matrix and $g_\pm(k),\ f(k)$ are given by

$$g_\pm(k) = \mp w_{L\pm}^+(k)(k\cdot\bar\sigma)w_{L\pm}(k) = \mp w_{R\pm}^+(k)(k\cdot\sigma)w_{R\pm}(k) = \mp\mathrm{sign}(k_0 \mp k_3)|k^2|,$$
$$f(k) = w_{L+}^+(-k)\,w_{R-}(k) = w_{L-}^+(k)\,w_{R+}(-k) = i(k_0 - k_3)^{1/2}(k_0 + k_3)^{1/2^*}. \tag{24}$$

As it turns out, their general expressions are not needed, because to compute the spectrum we shall use an anti-commutation relations preserving unitary transformation that reduces the DK-SM case to the SM-4 case, where diagonalization is trivial. Here we assumed that the usual SM-4 global $U(4)$ flavor rotation has been made to transform $R$ into its diagonal form $R_d$ in (14).

For the DK-SM case such rotation is not possible and we need to diagonalize two $16 \times 16$ blocks of the $32 \times 32$ matrix in (20). The DK-SM analog of (23) is

$$\mathcal{S}_1(k) = \begin{pmatrix} g_-\Gamma & M_{c+}^+ & 0 & M_{d+}^+ \\ M_{c+} & g_+\Gamma & -M_{d+}^+ & 0 \\ 0 & -M_{d+} & -g_-\Gamma & M_{c+}^* \\ M_{d+} & 0 & M_{c+}^T & -g_+\Gamma \end{pmatrix}, \quad \mathcal{S}_2(k) = \begin{pmatrix} g_-\Gamma & M_{c-}^+ & 0 & M_{d-}^+ \\ M_{c-} & g_+\Gamma & -M_{d-}^+ & 0 \\ 0 & -M_{d-} & -g_-\Gamma & M_{c-}^* \\ M_{d-} & 0 & M_{c-}^T & -g_+\Gamma \end{pmatrix}, \quad (25)$$

where the diagonal matrices $M_{d\pm}$ are given by

$$M_{d+} = \begin{pmatrix} -fC & 0 \\ 0 & f^*C \end{pmatrix}, \quad M_{d-} = \begin{pmatrix} -f^*C & 0 \\ 0 & fC \end{pmatrix}, \quad C = \begin{pmatrix} ch\lambda_1 & 0 \\ 0 & ch\lambda_2 \end{pmatrix}, \quad (26)$$

and the cross-diagonal matrices $M_{c\pm}$ are given by

$$M_{c+} = \begin{pmatrix} 0 & f^*S \\ -fS & 0 \end{pmatrix}, \quad M_{c-} = \begin{pmatrix} 0 & fS \\ -f^*S & 0 \end{pmatrix}, \quad S = \begin{pmatrix} sh\lambda_1 & 0 \\ 0 & sh\lambda_2 \end{pmatrix}. \quad (27)$$

Here we assumed that the $U(2) \times U(2)$ flavor rotations on the quark fields $Q_i^A, u^A, d^A$ have been carried out so that mass matrices $\mathcal{M}_{u,d}$ in (10) have been reduced to their middle factors in (12).

We are now ready to compute the spectrum of the DK-SM. Recall that the action for the free massive Dirac spinor with mass $m$ vanishes on-shell. Therefore, its off-shell action density vanishes whenever $k^2 = m^2$ and we can identify the spectrum of the free Dirac field with the set of zeroes of the characteristic equation derived for the $32 \times 32$ matrix in (20), which we shall call the spectral zero set. This allows us to compute the spectrum of the DK-SM, because as we shall prove below its spectral zero set is identical with that of SM-4 for a particular choice of masses of the SM-4 multiplet.

To prove this we shall show that the characteristic equations for the eigenvalue problems for the SM-4 and DK-SM are identical and thus have identical spectra. From this we will conclude that there exists a $32 \times 32$ anti-commutation relations preserving unitary transformation in the Grassmann space that transforms the DK-SM bilinear form in (20) into its SM-4 analog. In the end, despite the more complicated form of (25-27), the spectrum of DK-SM obeys the same rule as in the SM-4 case: it is determined by the set of eigenvalues of the middle factor $R$ in the Cartan decomposition (12) of its mass matrix.

Trying to derive analytically characteristic equations for arbitrary $16 \times 16$ matrices is a hopeless task. However, after some simplifications our problem can be reduced to dealing with a set of $4 \times 4$ matrices and one can write down the characteristic equations for both Dirac and DK cases in closed form.

First, we note that diagonalization of $\mathcal{S}_1(k)$ automatically leads to diagonalization of $\mathcal{S}_2(k)$. The next step is to realize that, since matrices $C, S$ in (27, 28) are diagonal, we can reduce the $16 \times 16$ problem to two $8 \times 8$ problems by taking $C = ch\lambda$, $S = sh\lambda$, $\lambda = \lambda_1, \lambda_2$. Lastly, inspection of the $\mathcal{S}_1(k)$ matrices in the SM-4 and DK-SM case reveals that these matrices become block-diagonal if we rename rows and columns. Thus at first sight analytically intractable eigenvalue problem effectively reduces to a set of tractable $4 \times 4$ problems. In the SM-4 case we obtain the two relevant $4 \times 4$ matrices

$$\mathcal{S}_1(k) = \begin{pmatrix} g_- & 0 & 0 & -f^* m^A \\ 0 & g_+ & f^* m^A & 0 \\ 0 & f m^A & -g_- & 0 \\ -f m^A & 0 & 0 & -g_+ \end{pmatrix}, \quad \mathcal{S}_2(k) = \begin{pmatrix} g_- & 0 & 0 & -f m^A \\ 0 & g_+ & f m^A & 0 \\ 0 & f^* m^A & -g_- & 0 \\ -f^* m^A & 0 & 0 & -g_+ \end{pmatrix}, \quad (28)$$

where $m^A = m e^{\pm \lambda_i}$, $m = m_{u,d}$, $i = 1, 2$, is the mass of one of the eight SM-4 quarks.

For the DK-SM case we obtain $4 \times 4$ block-diagonal form for $\mathcal{S}_\alpha(k)$ after writing (25) in $8 \times 8$ matrix form with each entry in the matrix containing $2 \times 2$ matrices that mix $A = 1,2$ and, separately, $A = 3,4$ elements only among themselves. Then it becomes apparent that in these $8 \times 8$ matrices elements $(1), (4), (6), (7)$ of the $C_1(k)$ vector form one 4-dimensional invariant subspace, while elements $(2), (3), (5), (8)$ form an orthogonal 4-dimensional invariant subspace. This allows us to write the characteristic equation for the $8 \times 8$ matrix as a product of two equations for two $4 \times 4$ matrices $\mathcal{S}_1^{(\pm)}(k)$ that are given by

$$\mathcal{S}_1^{(+)}(k) = \begin{pmatrix} g_- & -f^* s & 0 & -f^* c \\ -f s & -g_+ & -f c & 0 \\ 0 & -f^* c & g_- & -f^* s \\ -f c & 0 & -f s & -g_+ \end{pmatrix}, \quad \mathcal{S}_1^{(-)}(k) = \begin{pmatrix} -g_- & f s & 0 & f c \\ f^* s & g_+ & f^* c & 0 \\ 0 & f c & -g_- & f s \\ f^* c & 0 & f^* s & g_+ \end{pmatrix}, \quad (29)$$

where now $c = m\, ch\lambda, s = m\, sh\lambda, \lambda = \lambda_{1,2}$. Note that $\mathcal{S}_1^{(-)}(g_\pm, f) = \mathcal{S}_1^{(+)}(-g_\pm, -f^*)$.

We can now write the characteristic equation for the full $32 \times 32$ DK-SM and SM-4 problems as products of characteristic equations for eight $4 \times 4$ problems. Computing appropriate determinants, for the SM-4 case we obtain the same characteristic equation for $\mathcal{S}_1(k)$ and $\mathcal{S}_2(k)$, which results in the polynomial of degree 32

$$\prod_{A=1,\ldots,4} \left( (\rho + g_+)(\rho - g_-) - (m^A)^2 |f|^2 \right)^2 \left( (\rho + g_-)(\rho - g_+) - (m^A)^2 |f|^2 \right)^2 = 0, \quad (30)$$

where $\rho = \rho(k_0, k^2)$ is the spectral parameter. Using (24) it is easy to verify that $\rho(k_0, k^2 = (m^A)^2) = 0$ is a root of (30) for $A = 1,\ldots,4$ and that for $k^2 < 0$ the roots of (30) satisfy $\rho(k_0, k^2) > 0$. Analogous derivation for $\mathcal{S}_1^{(+)}(k)$ and $\mathcal{S}_1^{(-)}(k)$ in the DK-SM case results in

$$\left( (\rho + g_+)(\rho - g_-) - \mu_+^2 |f|^2 \right) \left( (\rho + g_+)(\rho - g_-) - \mu_-^2 |f|^2 \right) = 0, \quad (31)$$

$$\left((\rho+g_-)(\rho-g_+)-\mu_+^2|f|^2\right)\left((\rho+g_-)(\rho-g_+)-\mu_-^2|f|^2\right)=0, \tag{32}$$

respectively, where $\mu_\pm^2 = m^2 e^{\pm 2\lambda}$, $\lambda = \lambda_1, \lambda_2$. Combining (31, 32) with similar equations for $\mathcal{S}_2^{(+)}(k)$ and $\mathcal{S}_2^{(-)}(k)$ we conclude that the characteristic equation for the DK-SM case is identical to that of the SM-4 case, provided we take its masses proportional to the spectral set in (14). This implies that there exists a unitary Bogoliubov transformation that preserves the anti-commutation relations and transforms the DK-SM into a SM-4. In other words, the DK-SM theory with its Fock space defined through (15, 16) is equivalent to an SM-4 theory with the standard Dirac one-particle state operators obtained via Bogoliubov rotation of the DK-SM creation and annihilation operators.

Including the completely analogous down quark results we obtain that the quark sector spectrum of the DK-SM consists of eight masses given by the set

$$\{m_{u,d}^A\} = \left(m_{u,d} e^{-\lambda_1}, m_{u,d} e^{-\lambda_2}, m_{u,d} e^{+\lambda_1}, m_{u,d} e^{+\lambda_2}\right). \tag{33}$$

The eight quark masses are not independent. For the up and, separately, for the down quarks there exist a mass relation that follows from (33). For bare masses we obtain

$$m_u^1 m_u^3 = m_u^2 m_u^4, \tag{34}$$

$$m_d^1 m_d^3 = m_d^2 m_d^4. \tag{35}$$

Exactly the same analysis applies to the lepton sector of the DK-SM with Dirac neutrinos. Therefore, we obtain that in such a case the leptonic mass spectrum of the DK-SM and the tree level mass relations are given by

$$\{m_{\nu,e}^A\} = \left(m_{\nu,e} e^{-\lambda_3}, m_{\nu,e} e^{-\lambda_4}, m_{\nu,e} e^{+\lambda_3}, m_{\nu,e} e^{+\lambda_4}\right), \tag{36}$$

$$m_\nu^1 m_\nu^3 = m_\nu^2 m_\nu^4, \tag{37}$$

$$m_e^1 m_e^3 = m_e^2 m_e^4, \tag{38}$$

where $\lambda_3, \lambda_4$, $\lambda_3 > \lambda_4$, are some non-negative real numbers and $m_{\nu,e}$ are the two leptonic mass scales.

## 4. Experimental Constraints on Mass Spectrum and Mixing

We now compare DK-SM mass parameters with the experimental data. Note that, since $e^{+\lambda_1} > e^{+\lambda_2}$, according to the SM scheme of generation numbering in the order of increasing mass, in our formalism the third member of the DK-SM multiplet with $A = 3$ has the highest mass and, therefore, represents the fourth generation of the SM-4. The $A = 4$ member corresponds to the observed third generation of the SM. From now on we shall switch the numbering of the states with $A = 3, 4$ so that spectrum is ordered as in the SM.

Our mass predictions necessarily suffer from an inconsistency, caused by using the experimental mass values corresponding to renormalized masses obtained from unrelated bare masses. The correct approach is to bring all mass parameters to the same energy scale using the renormalization group analysis based on DK-SM. Still, even tree level predictions can give us some feeling about the degree of the agreement between our model and the experiment and thus provide a test for our model.

We begin with the lepton mass spectrum. The masses of the first three generations of the charged leptons are known with high accuracy. However, considering the potentially large differences caused by differing energy scales used in mass computations, we shall quote the predicted values with two-digit accuracy only. Using (38) and the charged lepton mass values from [17] we obtain for the fourth charged lepton, provisionally called the $\kappa$ (kappa)-lepton,

$$m_{e_4} = \frac{m_\mu m_\tau}{m_e} \cong 370 \, GeV \,. \tag{39}$$

For neutrinos only the differences in squared masses are known. From [17] we obtain $\Delta m_{12}^2 \cong 7.6 \cdot 10^{-5} \, eV^2$, $\Delta m_{32}^2 \cong 2.4 \cdot 10^{-3} \, eV^2$. We now note that unless $m_\nu^1 << m_\nu^2$, mass relation (37) would imply the existence of the fourth light neutrino, which is ruled out by the experiment [17]. We, therefore, have to assume that $m_{\nu_e} << m_{\nu_\mu}$. Therefore, our model selects from possible mass identifications the normal hierarchy for neutrino masses with $(m_{\nu_\mu})^2 = \Delta m_{12}^2 \cong 7.6 \cdot 10^{-5} \, eV^2$, $(m_{\nu_\tau})^2 = \Delta m_{32}^2 \cong 2.4 \cdot 10^{-3} \, eV^2$. Hence, we obtain for neutrino masses for the second and the third generations $m_{\nu_\mu} \cong 8.7 \cdot 10^{-3} \, eV$, $m_{\nu_\tau} \cong 4.9 \cdot 10^{-2} \, eV$. At the same time we obtain a see-saw-like mass relation for the electron neutrino and $\kappa$ - neutrino

$$m_{\nu_4} m_{\nu_e} = m_{\nu_\mu} m_{\nu_\tau} \cong 4.2 \cdot 10^{-22} \, GeV^2 \,. \tag{40}$$

Since experimentally $m_{\nu_4} > 100 \, GeV$ [17] [2], we obtain $m_{\nu_e} < 10^{-24} \, GeV$. Note that throughout the paper we assumed that neutrinos are Dirac particles.

We now turn to quark mass spectrum. Free quarks have never been observed and for up or down quark mass one does not have direct experimental measurements. Most reliably these masses are computed using lattice simulations that use the SM. A more consistent approach would be to use DK-SM, which would not use the $3 \times 3$ unitarity constraints. Even taking the SM calculations as applicable for our purposes there is a large uncertainty in the computed masses of the up and down quarks. The choice of the subtraction scheme used to determine $m_s$ also causes a significant variation in the mass of the strange quark. With all these reservations using data in [17] we obtain that

$$m_{t'} \approx (60 - 140) \, TeV \,, \tag{41}$$

$$m_{b'} \approx (60 - 150) \, GeV \,. \tag{42}$$

The range for $m_{t'}$ is too high for $t'$ detection at LHC. At the same time the tree level range for $m_{b'}$ is too low. The most current lower bound on mass of $m_{b'}(m_{t'})$, derived using the SM, is $385(335) \, GeV$ [18]. If the range in (42) cannot be brought up by using consistent DK-SM calculations of the radiative corrections or other effects then our model would have a serious problem. The mass values for the first three generations, the mass estimates of the 4[th] generation, and related parameters are summarized in Fig. 1. We observe that $\lambda_i$ and $m$ parameters for down quarks and charged leptons differ by a relatively small factor. On the other hand, neutrinos stand out with very small $m$, while the electron neutrino and the fourth neutrino have very large $|\lambda_1| \approx 30$. If we assume that $|\lambda_1| \approx 10$ as is for other lepto-quarks then the 4$^{th}$ neutrino mass should lie below $1\ KeV$.

---

[2] Despite this bound the possibility of existence of sterile neutrinos in << 100 GeV mass range is under discussion. See for example [21].

The precision EW data and the SM impose a well-known constraint on the masses of the 4$^{th}$ generation quarks in SM-4. At one-loop level the quark mass difference is given [7, 8, 11] by $m_{t'} - m_{b'} \approx (1 + (1/5)\ln(m_H/115 GeV)) \times 55\ GeV$. This constraint is compatible with (34, 35) if we put $m_H \cong \infty$, which is another way to say that the Higgs field decouples from the fermionic sector of our theory. However, in order for our model to become phenomenologically viable, the large difference $m_{t'} - m_{b'} \approx 10^3\ GeV$ needs to be reconciled with all EW data. We shall discuss this and related issues in the forthcoming publication.

|  | $v$ | $e$ | $d$ | $u$ |
|---|---|---|---|---|
| $\lambda_1$ | $> 29$ | $9.0$ | $4.6 - 5.3$ | $8.3 - 9.1$ |
| $\lambda_2$ | $0.82 - 0.93$ | $1.4$ | $1.5 - 2.3$ | $1.7 - 2.5$ |
| $m$ | $(2.0 - 2.1) \cdot 10^{-11}$ | $0.43$ | $0.57 - 0.79$ | $14 - 15$ |
| $m_1$ | $< 10^{-24}$ | $5.1 \cdot 10^{-4}$ | $(4.1 - 5.8) \cdot 10^{-3}$ | $(1.7 - 3.3) \cdot 10^{-3}$ |
| $m_2$ | $(8.6 - 8.8) \cdot 10^{-12}$ | $0.11$ | $0.08 - 0.13$ | $1.3$ |
| $m_3$ | $(48 - 50) \cdot 10^{-12}$ | $1.8$ | $(4.1 - 4.8)$ | $170$ |
| $m_4$ | $> 100$ | $370$ | $(60 - 150)$ | $(60 - 140) \cdot 10^3$ |

Fig. 1. Masses (GeV) and related parameters in the DK-SM. The underlined mass values are DK-SM estimates. The remaining mass values, quoted with two digit precision, are from [17]. The spread for the strange quark mass is the combined spread for MS and 1S subtraction schemes.

We now briefly describe mixing in the DK-SM. In Section 3 we proved that there exists a Bogoliubov transformation in the phase space that transforms the off-shell free action of the DK-SM model into the free action of SM-4 model with the same mass spectrum. We note now that, in principle, the same can be done for the interacting theory. That is, there exists a Bogoliubov transformation in the phase space that preserves the anti-commutation relations, redefines one-particle states, and transforms the full action (19) of the DK-SM into action of the SM-4. This conjecture is reasonable, because the gauging procedure both for Dirac and DK spinors is essentially unique. It holds true for the degenerate case when $\lambda_i = \lambda_j$ within DK multiplet.

One consequence of the conjecture is the form of the $4 \times 4$ unitary mixing matrix of the corresponding SM-4. The quark mixing matrix is defined as $W^{DK} = W_L^u W_L^{d+}$. Since according to our conjecture $W_L^{u,d} = W_{14,23}^{u,d} W_{12,34}^{u,d}$, where index $ij, kl$ denotes $U(2)$ mixing of generations $ij$ and, separately, generations $kl$, we obtain that the full mixing matrix must have the form

$$W^{DK} = W_{14,23}^u \widetilde{W}_{12,34}^{ud} W_{14,23}^d \equiv W_B^u W_P W_B^d \tag{43}$$

where $W_B^{u,d}$, $W_P$ belong to two different $U(2) \times U(2)$ subgroups of flavor $U(4)$. Similar decomposition exists for the leptonic sector. $W_B^{u,d}$, called the Bogoliubov or b-mixing matrices, mix only $A = 1, 4$ and separately $A = 2, 3$ states and have the texture

$$W_B^{u,b} = \begin{pmatrix} \times & 0 & 0 & \times \\ 0 & \times & \times & 0 \\ 0 & \times & \times & 0 \\ \times & 0 & 0 & \times \end{pmatrix}, \quad W_B^{u,d} \in U(2) \times U(2). \tag{44}$$

At the same time $W_P$, called the premixing or p-matrix, mixes only $A = 1, 2$ and separately $A = 3, 4$ states and has the texture

$$W_P = \begin{pmatrix} \times & \times & 0 & 0 \\ \times & \times & 0 & 0 \\ 0 & 0 & \times & \times \\ 0 & 0 & \times & \times \end{pmatrix}, \quad W_P \in U(2) \times U(2), \tag{45}$$

where $\times$ denotes non-zero entries.

Decomposition of $W_P$ into direct product of two $U(2)$ factors follows from its definition. For $W_B^{u,d}$ it follows from the factorization of the characteristic equations (31, 32). It is important to note that the order of the factors in (43) is fixed. This is because we performed the removal of the $U(2) \times U(2)$ factors in (12) first and only then applied the diagonalizing Bogoliubov transformation.

We shall postpone detailed discussion of the properties of $W^{DK}$ till a follow-up publication. However, as an example, we note that $W^{DK}$ in (47) reproduces the known near-tri-bimaximal mixing texture of the leptonic sector [17, 19, 20]. If we set $W_B^u = I$ and for the remaining factors take

$$W_B = \begin{pmatrix} c_{14} & 0 & 0 & s_{14} \\ 0 & c_{23} & s_{23} & 0 \\ 0 & -s_{23} & c_{23} & 0 \\ -s_{14} & 0 & 0 & c_{14} \end{pmatrix}, \quad V_P = \begin{pmatrix} c_{12} & s_{12} & 0 & 0 \\ -s_{12} & c_{12} & 0 & 0 \\ 0 & 0 & c_{34} & s_{34} \\ 0 & 0 & -s_{34} & c_{34} \end{pmatrix}, \quad (46)$$

where $W_B^e \equiv W_B$, $s_{AB} \equiv \sin\theta_{AB}$, *etc*, we obtain for the upper $3\times 3$ block

$$W_{SM}^{DK} = \begin{pmatrix} c_{14}c_{12} & c_{14}s_{12} & -s_{14}s_{34} \\ -c_{23}s_{12} & c_{23}c_{12} & s_{23}c_{34} \\ s_{23}s_{12} & -s_{23}c_{12} & c_{23}c_{34} \end{pmatrix}. \quad (47)$$

For $c_{14} = 1$, $s_{14} = 0$, $c_{23} = \sqrt{1/2}$, $s_{23} = -\sqrt{1/2}$, we obtain from (47) a near tri-bimaximal form for the PMNS mixing matrix

$$W_{SM}^{DK} = U_{PMNS} = \begin{pmatrix} c_{12} & s_{12} & 0 \\ -s_{12}/\sqrt{2} & c_{12}/\sqrt{2} & -c_{34}/\sqrt{2} \\ -s_{12}/\sqrt{2} & c_{12}/\sqrt{2} & c_{34}/\sqrt{2} \end{pmatrix}. \quad (48)$$

Non-zero $s_{14}s_{34}$ can also accommodate a non-vanishing $(e3)$ element of $U_{PMNS}$. Choosing $c_{14} \cong 1$, $|s_{14}| \ll 1$, $c_{23} \cong 1$, $|s_{23}| \ll 1$ reproduces the texture of the absolute values of $V_{CKM}$ in the quark sector, where mixing of the third generation is suppressed in comparison to the first two.

## 5. Summary

We established that the use of Dirac-Kähler spinors instead of Dirac spinors leads to constraints on mass spectrum and mixing of the SM-4 extension. The spectrum of the DK-SM obeys mass relations that can be used to estimate masses of the 4$^{th}$ generation quarks, the mass of the fourth charged lepton, to select normal hierarchy for neutrino masses, and to establish a see-saw-like relation between the electron and the 4$^{th}$ neutrinos. The tree-level estimate of the 4$^{th}$ charged lepton turns out to be very precise. The 4$^{th}$ generation bottom mass estimate appears to be too low and is ruled out by the experiment. Our mass estimates are given at tree level and definitive predictions have to include the radiative corrections computed within the framework of the DK-SM. Considering the relatively small differences in the $\lambda_i$ and $m$ parameters for most of the particles, we should expect that the correction should not be by a large factor. Therefore, we should expect that the mass of the fourth charged $\kappa$-lepton and of $b'$ should lie within the operational range of LHC.

Whether or not the DK-SM can serve as a phenomenologically viable extension of the SM still remains an open question. The large predicted difference in masses between $t'$ and $b'$ must be explained, in view of strong EW constraints coming from the oblique parameters [7, 8, 11, 22]. In addition to addressing the potential $m_{b'}$ problem, the known absolute values and phases of the CKM and PMNS mixing matrices must be computed within the framework of the DK-SM and agree with the known experimental values. We shall address these issues in another publication.

## Acknowledgement

I would like to thank the members of the Elementary Particle Theory Group at the IKTP, Dresden for interesting discussions.